\documentclass{article}
\pdfoutput=1

\usepackage{graphicx,comment}
\usepackage{amsmath}
\usepackage[a4paper]{geometry}

\title{A novel method for determining the phase-response curves of neurons based on minimizing spike-time prediction error}
\author{Ben Torben-Nielsen, Marylka Uusisaari, Klaus M. Stiefel}
\date{\today}

\begin{document}

\maketitle

\begin{abstract}
Regular firing neurons can be seen as oscillators. The phase-response
curve (PRC) describes how such neurons will respond to small excitatory perturbations.
Knowledge of the PRC is important as it is associated to the excitability
type of neurons and their capability to synchronize in networks. In
this work we present a novel method to estimate the PRC from experimental
data. We assume that continuous noise signal can be discretized into
independent perturbations at evenly spaced phases and predict the next spike based on these
independent perturbations. The difference between the predicted next spike made at every discretized phase and the actual next spike time is used as the error signal used to optimize the PRC. We
test our method on model data and experimentally obtained data and find
that the newly developed method is robust and reliable method for
the estimation of PRCs from experimental data.
\end{abstract}

\section{Introduction}

The phase response curve (PRC) is a property of oscillators that defines
how the phase of the next cycle is shifted by an incoming perturbation
at a particular phase. In the case of regular firing neurons, the
PRC describes how the timing of the next spike is shifted. The PRC
is of interest for several reasons. First, by using the PRC a prediction
can be made about the synchronizing properties of synoptically coupled
neurons. Nonnegative PRCs (i.e., type-I PRCs or monophasic PRCs) do
not allow for synchronization via excitatory connections while PRCs with a negative part (i.e.,
type-II PRCs or biphasic PRCs) allow for synchronization in networks
with excitatory coupling and small conduction delays \cite{hansel1995}.
Second, the PRC type indicates which bifurcation type leads to the
transition from non-spiking to spiking behavior \cite{izhikevich2007}.
Moreover, the PRC type correlates with the type of excitability of
a neuron \cite{hodgkin1952,marella2008}. As such, the PRC is an important
characterizing feature of regularly firing neurons.

The PRC can be easily understood when projecting the higher-dimensional
dynamics of a model of a neuron to a one-dimensional system, i.e.,
when a neuron is seen as an oscillator, with its state determined
by the phase $\phi$. The unperturbed state of the neuron is then
described as $\frac{{d\phi}}{{dt}}=\omega$, with $\omega=\frac{2\pi}{T}$
and $T$ the average inter-spike interval. With the introduction of
noise $x\left(\phi(t)\right)$ of small amplitude $\alpha$, the state
of the neuron is described as \begin{equation}
\frac{{d\phi}}{{dt}}=\omega+\alpha\cdot x\left({\phi(t)}\right)\cdot Z\left({\phi(t)}\right)\end{equation}
 where $Z\left({\phi(t)}\right)$ is the PRC. By using a shorthand
notation $\phi=\phi(t)$ we can describe the spike times of the regularly firing neuron
as \begin{equation}
{\bf {S}}=\omega+\alpha\cdot x\left(\phi\right)\cdot Z\left(\phi\right)\end{equation}
 Thus, the firing times of the neuron is determined by the period
$\omega$, the noise signal and the PRC $Z(\phi)$.

The PRC of a neuron is usually estimated from experimental data by the direct method in which small, brief current-pulses are injected at different phases and
the phase shifts of the next spike are recorded. By following the definition
of the PRC, the phase of the perturbation is plotted on the x-axis
and the resultant shift in normalized spike time (phase) on the y-axis.
In practice, neuron spike times display a large amount of jitter
and therefore it is necessary to measure the spike time shifts in
hundreds of inter-spike intervals at random phases to span all phases
and to obtain a reliable readout. 

An alternative method was proposed
by Izhikevich \cite{izhikevich2007}, where continuous current fluctuation was injected instead of brief pulses.  Subsequently, by summing
the effects of the injected fluctuations between two spikes as predicted
by a a candidate PRC, the time of the next spike is estimated. The
summed difference between the observed and predicted spike times
is then used as an error signal to optimize the PRC. In practice this
method does not converge within a reasonable number of optimization
rounds (e.g., 5000 iterations with a basic simplex optimization algorithm). The inability to converge is likely caused by (i) the loss of temporal information when the errors are summed, and (ii) the fact that many candidate PRCs can have the same error value. We thus developed an extension of this method which does
not sum over these two dimensions and instead uses an array of error signals that preserve the phase information to optimize the PRCs.

Our novel method, the STEP (Standardized Error Prediction) method, allows for an reliable
estimation of the PRC with relatively few spikes and at high noise levels.
The STEP method is a numerical method based on the assumption that the effects
of small perturbations on the PRCs are independent and treating them independently preserves temporal information.

\section{Methods}

\subsection{Experimental procedure}

\begin{figure}
\centering
\includegraphics[scale=1]{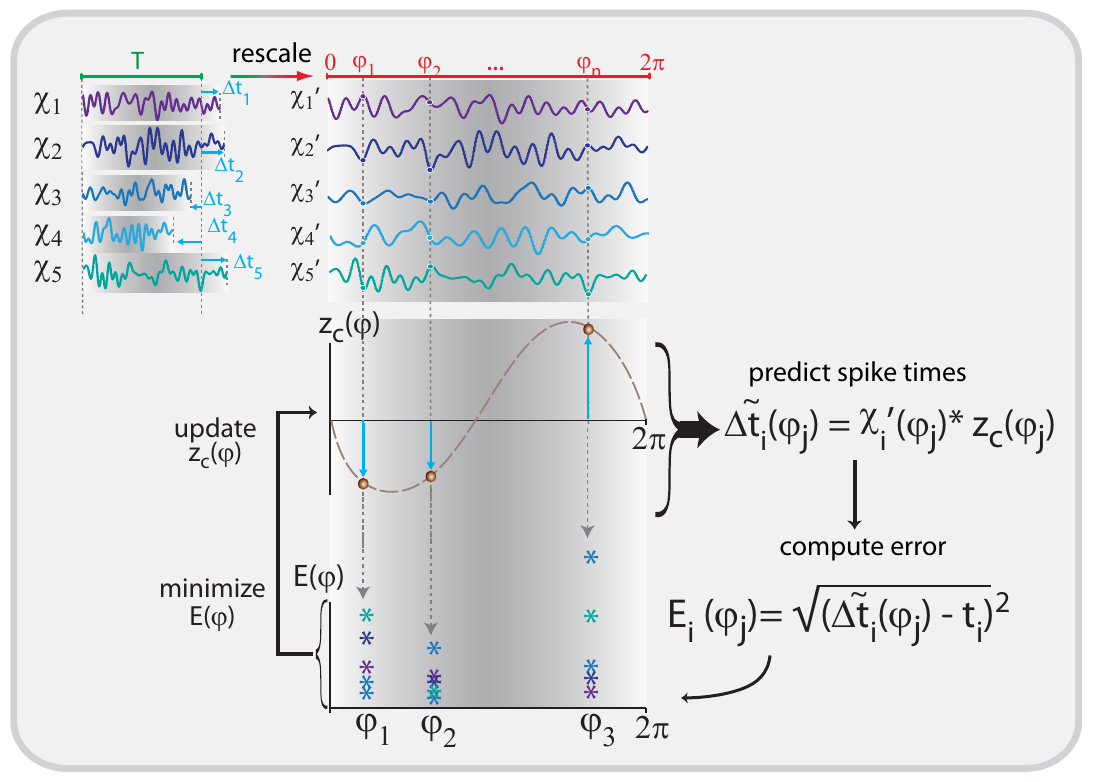}
\caption{Experimental procedure of the STEP method. The input fluctuation between two spikes ($x_n$) are scaled to the length of the average ISI. Then, at a predetermined number of `bins' in the normalized phase ($\{\phi_1,\phi_2,\ldots,\phi_n\}$), the next spike time is predicted according to a candidate PRC $z_c(\phi)$. The difference between the true next spike time and the estimated spike time is used as error-signal. Due to the binning of normalized phases, we get a temporally structured error signal for each spike at each binned phase which allows for optimization with least-squares.}
\label{fig:method}
\end{figure}

We consider a neuron that is brought to regular firing by a constant
current injection (step current $I_{s}$). When stabilized, the considered
neuron fires with a period of $\omega=\frac{2\pi}{T}$, where T is
the average interspike interval. A continuous fluction is imposed
on top of $I_{s}$ and the combined signal is defined as $x(t)$ (the total current
injected into the neuron). In the STEP method, the PRC is approximated
by a truncated Fourier series $z(\phi)\approx\sum\limits _{n=0}^{n=3}{{\bf {a}}\sin(n\phi)+{\bf {b}}\cos(n\phi)}$
where $z(\phi)$ is a candidate PRC of which ${\bf {a}}$ and ${\bf {b}}$
are subject to optimization. The optimization is guided by the distance
between predicted spike times $\hat{s}_{i}$ to the real spike times
$s_{i}$. We implicitly normalize the duration of the fluctuating
signal between subsequent spikes by equating this interval to $[0,2\pi]$
($s_{i}\equiv0$ and $s_{i+1}\equiv2\pi$). Then, we discretize the
period of the oscillation into N bins,$\left\{ \phi_{1},\phi_{2},\ldots,\phi_{n}\right\} $.
Subsequently, for every phase bin $\phi_{j}$, based on the fluctuating
signal $x(t)$, we compute the predicted next spike time $\hat{s}_{i,j}=s_{i-1}+x(t(\phi_{j}))\cdot z(\phi_{j})$
and formulate an error signal $E_{i,j}=\sqrt{\left(\hat{s}_{i,j}-s_{i}\right)^{2}}$.
Hence, for every spike we obtain a error signal for each phase $\phi_{j}$
and use $E_{i,j}$ with least-squares optimization to optimize the
${\bf {a}}$ and ${\bf {b}}$ of the truncated Fourier series. Figure~\ref{fig:method}
illustrates this procedure. (In all shown simulations, we use 100
bins.)

\subsection{Model neuron and experimental data}

We use three data sets to test our method. The first data set contains noise-free model data in which a single
compartmental model neuron (see below) is perturbed at different phases.
This set contains 128 perturbations evenly spaced over $[0,2\pi]$.
This data set was used as a benchmark to compare the stochastic simulations
to.

The second data set contains modeled data from the same single-compartmental
model but with an additionally injected fluctuating current. The fluctuations
are generated through a stationary Orstein-Uhlenbeck process around
a given mean value and parametrized by the reversion rate ($g=0.1$)
and 4 different volatility levels ($D={1e^{-4},5e^{-4},1e^{-5},5e^{-5}}$).
The advantage of the modeled data (with continuous fluctuations) is
that we know the excitability type with certainty because small perturbations
do not change the PRC type \cite{izhikevich2007}. 

The single compartment model neuron is a modification of the model
developed by Golomb and Amitai \cite{golomb1997}, as modified in
\cite{stiefel2009}. It uses a Hodgkin Huxley type formalism to model
neural spiking behavior\begin{equation}
C_{M}\frac{{dV}}{{dt}}=-m^{3}h\overline{g}_{Na}\left({V-E_{Na}}\right)-n\overline{g}_{KDR}\left({V-E_{K}}\right)-s\overline{g}_{Ks}\left({V-E_{K}}\right)-\overline{g}_{leak}\left({V-E_{leak}}\right)-I_{inj}\end{equation}
and $\frac{{dx}}{{dt}}=\tau\left(V\right)\left({x-x_{\infty}\left(V\right)}\right)$where
$V$ is the membrane potential, $\overline{g}_{x}$ the maximum conductance
for ion $x$ and$E_{x}$ the reversal potential for ion $x$. The
parameter values can be found \cite{golomb1997,stiefel2009}. By turning
the adaptation current on or off, this model switches between type-II
or type-I excitability \cite{stiefel2009}, respectively. The model
is simulated using NEURON \cite{carnevale2006} and simulates 100s
of neuronal time which results in approximately 950 spikes.

The third data set contains experimental data recorded from a layer
2/3 pyramidal cell of the mouse cortex with the whole cell patch-clamp
techniques in vitro. Standard patch-clamp techniques as in \cite{stiefel2009} were used. Membrane potential voltage
data and the injected fluctuations were digitized at 40 kHz. Two levels of fluctuation amplitudes (50 pA and 100pA) were tested and resulted in 655 and 647 spikes used; only spikes that satisfied $0.1 \times \widetilde{ISI} \leq ISI \leq 2  \times \widetilde{ISI}$ were used.

\section{Results}

\begin{figure}
\begin{center}
\begin{tabular}{cc}
\includegraphics[scale=0.3]{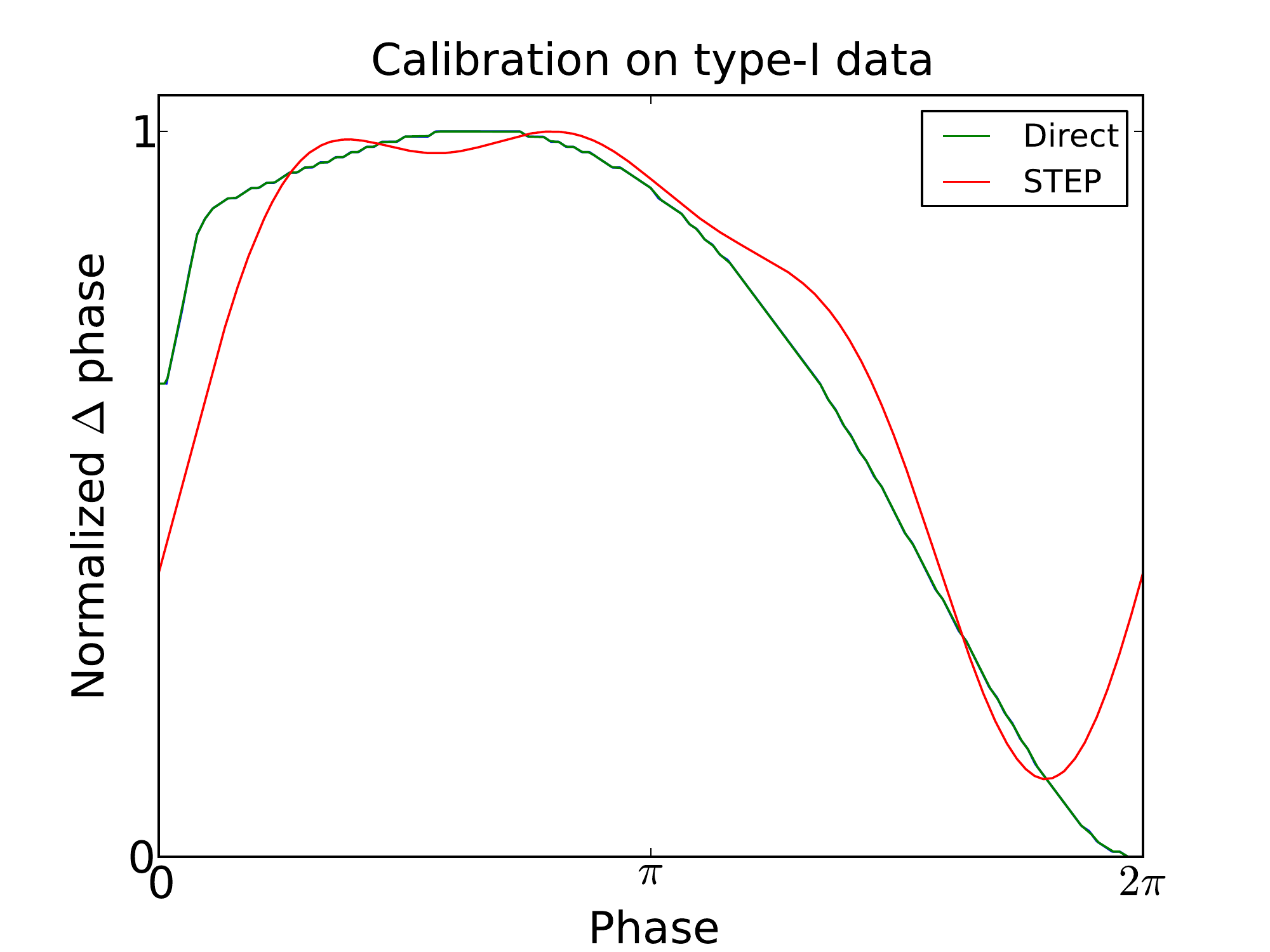} &
\includegraphics[scale=0.3]{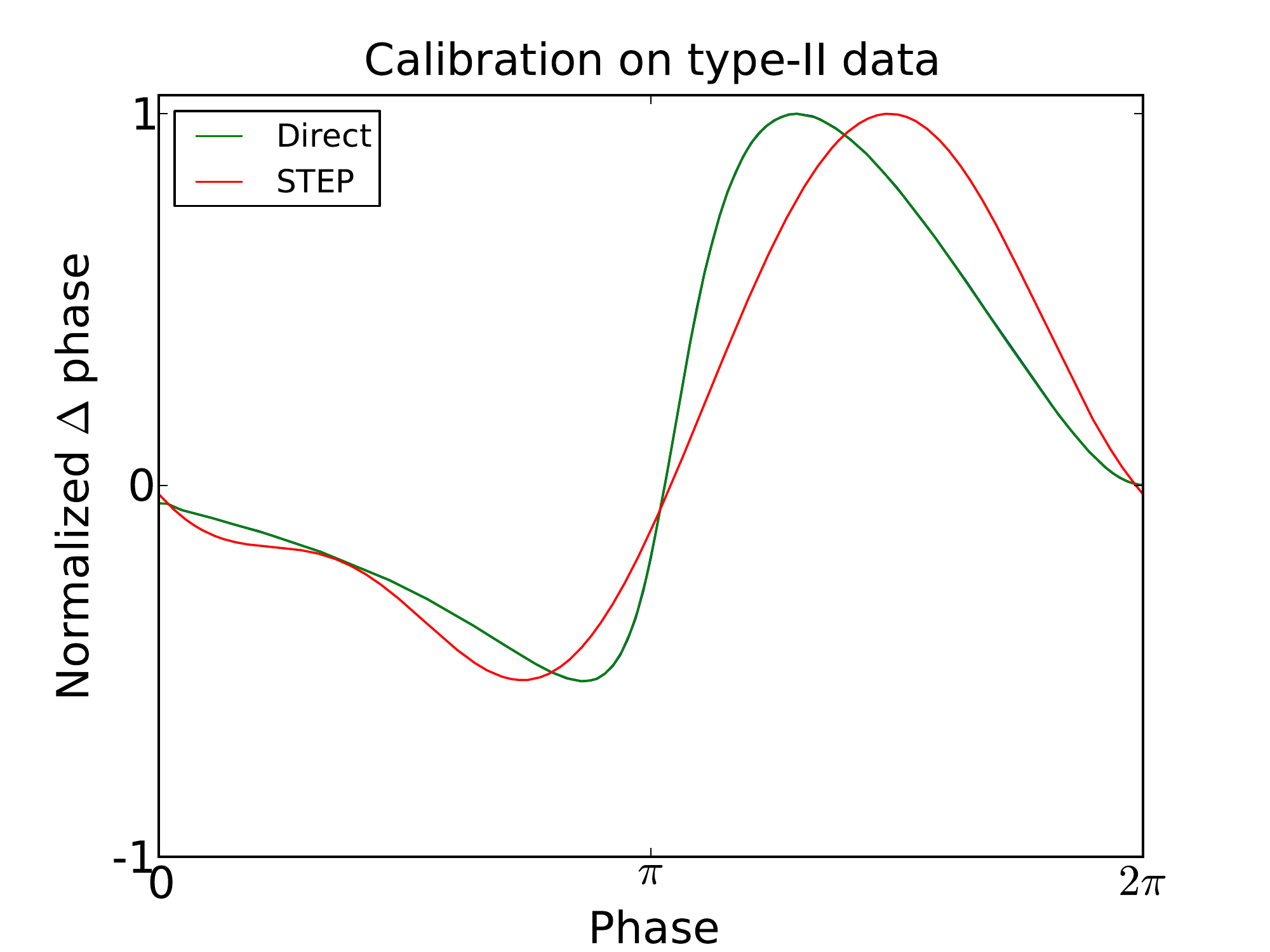}
\end{tabular}
\end{center}
\caption{Calibration of the STEP method. PRCs obtained with the STEP method are compared with those obtained from the same data by using the direct method.}
\label{fig:calibrated}
\end{figure}

To calibrate the data, we injected short pulses at different phases
and directly recorded the phase shift. Afterwards, we ran the STEP
method on the perturbation data and found that the resulting PRC matches
well with the directly determined PRC (Figure~\ref{fig:calibrated}).
The slight mismatch at the end of the period in type-I data and middle
of the type-II data is due to the short expansion (3rd order) of the
Fourier series. Longer expansion tend to provide a better fit but
are prone to over-fitting (not shown).

\begin{figure}
\begin{center}
\begin{tabular}{c}
\includegraphics[scale=0.3]{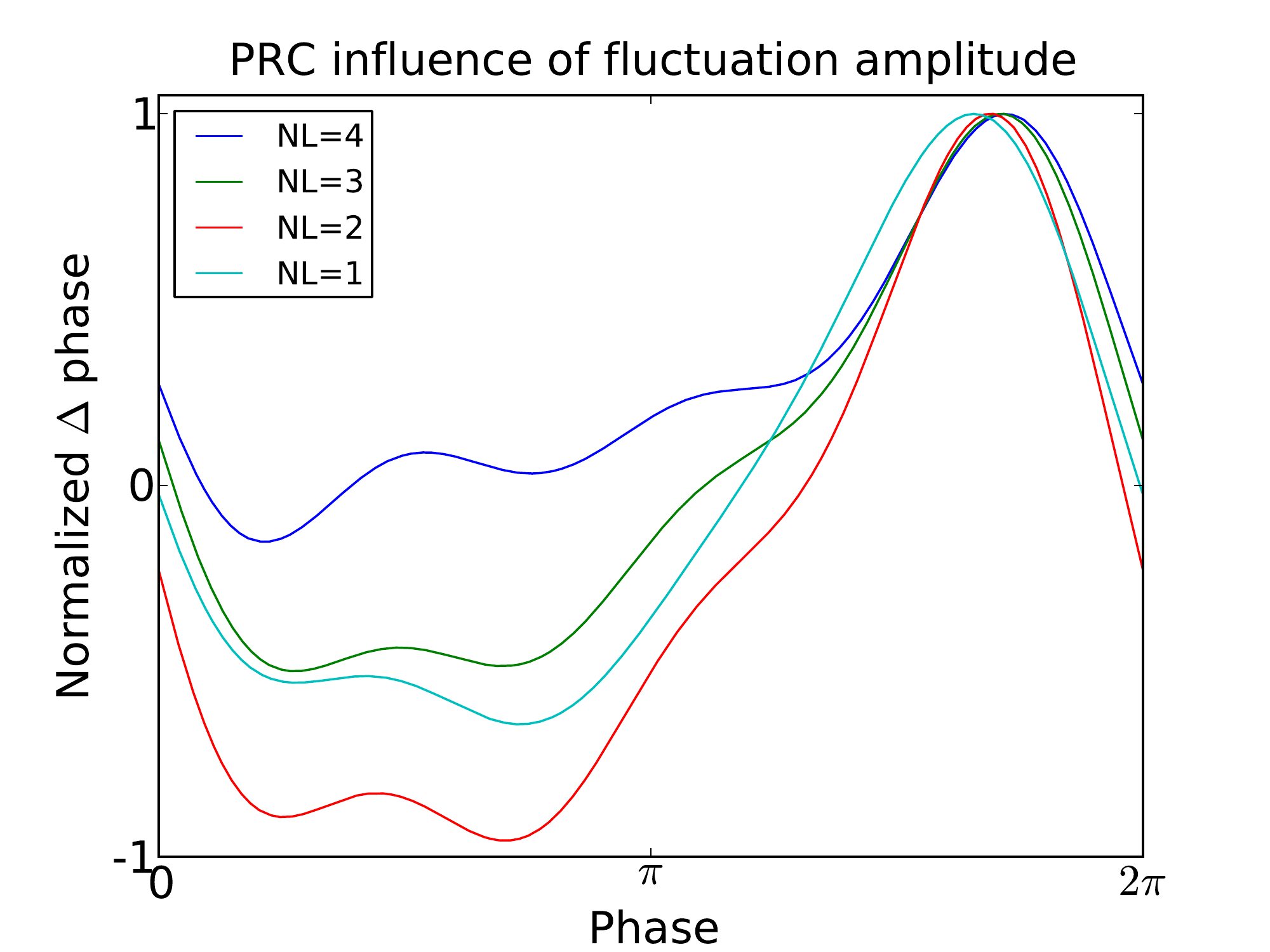}
\includegraphics[scale=0.3]{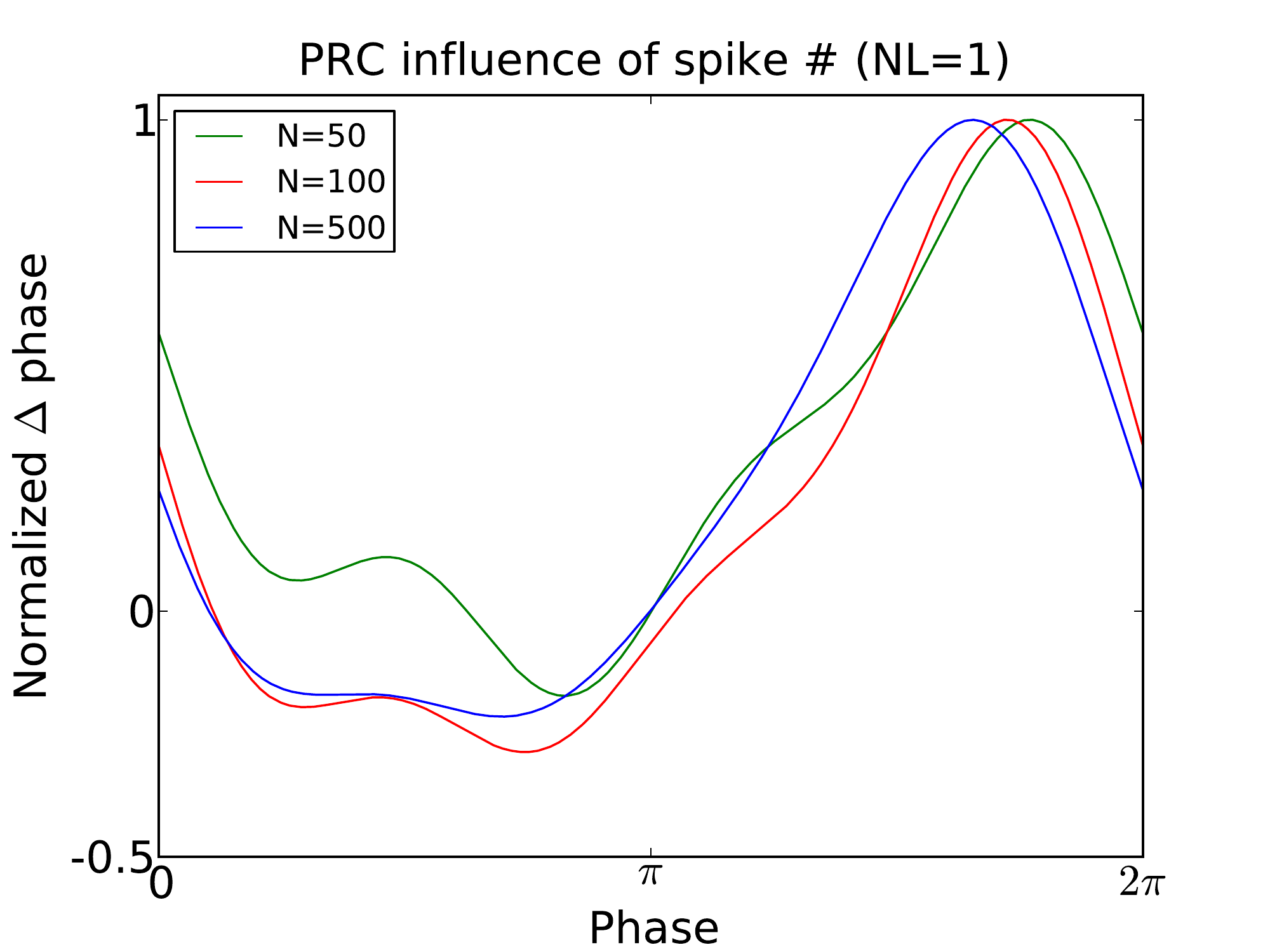}
\end{tabular}
\end{center}
\caption{Performance of the STEP method with in terms of fluctuation level
and number of spikes. Top panel: the injected fluctuation signals.
Bottom left: the PRC generated by our method for the four different
noise levels. All PRCs are of type-II as they should be. Bottom right:
the estimated PRC based on 50, 100 and 500 spikes (for the model data
at fluctuation level 1). The PRCs indicate are type-II and from 100
spikes and more the generated PRC converges. }
\label{fig:noise}
\end{figure}

Having established that STEP method works well with simple perturbation,
we tested our method on the more realistic case in which a continuous
fluctuation signal perturbs the neuron. The PRCs for the four different
noise levels Figure~\ref{fig:noise} (bottom left) and indicate that
at all noise levels the PRC correctly displays type-II characteristics
 (a substantial negative part). 
 
 An important feature of any method
to estimate the PRC is its sensitivity to the number of spikes because
large numbers of spikes are hard to obtain experimentally. For this reason, we tested
our method with 50, 100 and 500 spikes (Figure~\ref{fig:noise},
right). We observe that the STEP algorithm provides a fair estimation with
a minimum of 100 spikes. This number of required spikes is much lower
than most other methods, for instance the STA method \cite{ermentrout2007} is shown to work with 7000 spikes, Galan's method \cite{galan2005} with 480 spikes and the WSTA method \cite{ota2009} with `a few hundred' spikes.

\begin{figure}
\begin{center}
\begin{tabular}{cc}
\includegraphics[scale=0.3]{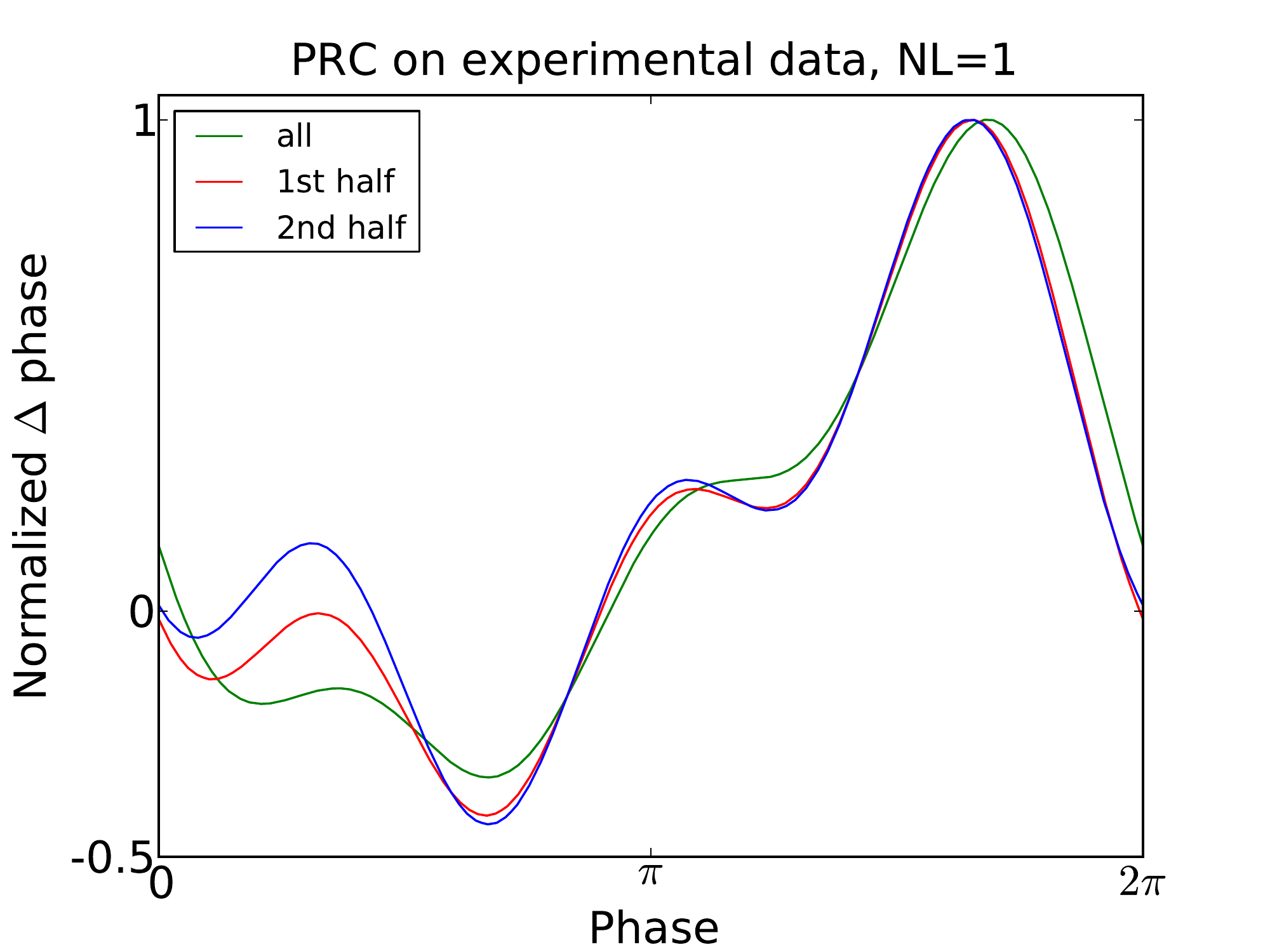} &
\includegraphics[scale=0.3]{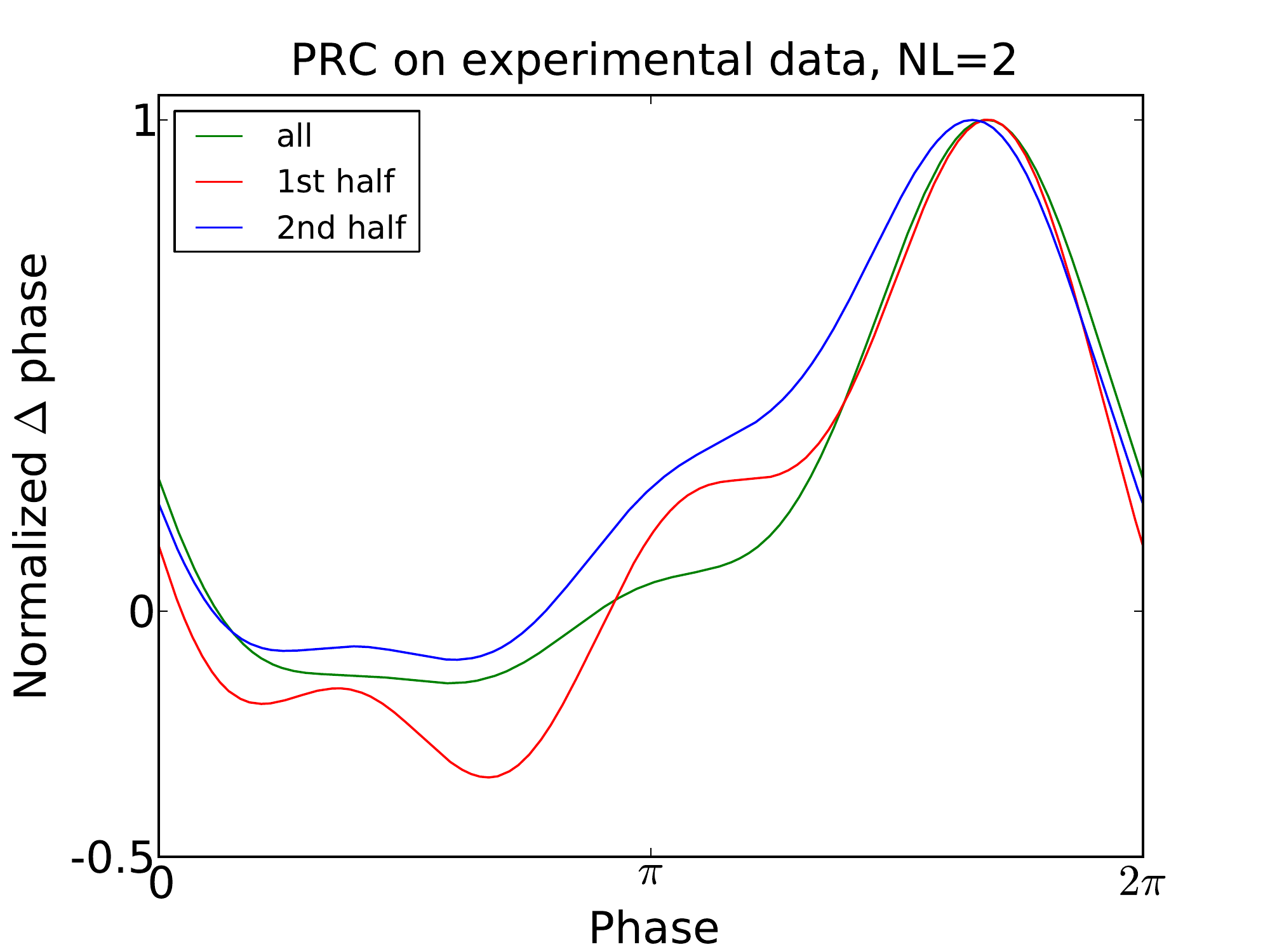}
\end{tabular}
\end{center}
\caption{PRC generated by the STEP method on experimental data. To investigate the reliability of the produced PRC we generated two additional PRCs each with only half the available spiking data. Since all three PRCs are similar (all 3 per noise level, and all 6 over both noise levels), the resulting PRCs can be considered reliable representation of the spiking data.}
\label{fig:experimental}
\end{figure}

Finally, we tested our method on experimental data obtained from mouse
layer 2/3 pyramidal neurons in the visual cortex. Unfortunately, the
PRC types for such neurons are not known as there is
experimental evidence that Layer 2/3 neurons can possess both types
of PRC \cite{stiefel2009}. To give an indication that our method works well with experimental data we used a strategy outlined in \cite{galan2005} in which we compare the PRC produced with all spike data to two PRCs produced with only half of all spiking data. If all the resulting PRCs resemble each other, the PRCs are unlikely to contain under- or over-fitting artifacts and can therefore be considered representative for the provided data. Figure~\ref{fig:experimental} shows these results.

We may state that the STEP method is a reliable method to estimate
the PRC from experimental data. The advantages of the STEP method are the low number of
required spikes and the robustness at higher noise levels.

The authors thank Drs. Stijn Vanderlooy, Yasuhiro Tsubo and Adam Ponzi for fruitful discussions.

\bibliographystyle{nar}
\bibliography{}

\end{document}